\documentclass[conference]{IEEEtran}
\IEEEoverridecommandlockouts
\usepackage{cite}
\usepackage{amsmath,amssymb,amsfonts}
\usepackage{graphicx}
\usepackage{textcomp}
\usepackage{xcolor}
\usepackage{algorithm}
\usepackage{algpseudocode}
\usepackage{caption} 
\usepackage{url}
\newcommand{\pluseq}{\mathrel{+}=}
\usepackage[colorinlistoftodos]{todonotes}

\begin{document}

\title{The Signaler-Responder Game: Learning to Communicate using Thompson Sampling}

\author{\IEEEauthorblockN{Radhika Bhuckory and Bhaskar Krishnamachari}
\IEEEauthorblockA{\textit{Ming Hsieh Dept. of Electrical and Computer Engineering} \\ \textit{Viterbi School of Engineering}\\
\textit{University of Southern California}\\
Los Angeles, USA \\
\{bhuckory, bkrishna\}@usc.edu}
}

\maketitle

\begin{abstract}
We are interested in studying how heterogeneous agents can learn to communicate and cooperate with each other without being explicitly pre-programmed to do so. Motivated by this goal, we present and analyze a distributed solution to a two-player signaler-responder game which is defined as follows. The signaler agent has a random, exogenous need and can choose from four different strategies: never signal, always signal, signal when need, and signal when no need. The responder agent can choose to either ignore or respond to the signal. We define a reward to both agents when they cooperate to satisfy the signaler's need, and costs associated with communication, response and unmet needs. We identify pure Nash equilibria of the game and the conditions under which they occur. As a solution for this game, we propose two new distributed Bayesian learning algorithms, one for each agent, based on the classic Thompson Sampling policy for multi-armed bandits. These algorithms allow both agents to update beliefs about both the exogenous need and the behavior of the other agent and optimize their own expected reward. We show that by using these policies, the agents are able to intelligently adapt their strategies over multiple iterations to attain efficient, reward-maximizing equilibria under different settings, communicating and cooperating when it is rewarding to do so, and not communicating or cooperating when it is too expensive.

\end{abstract}

\begin{IEEEkeywords}
Thompson Sampling, Learning to Communicate, Game Theory, Nash Equilibrium, Multi-Agent Systems, Cognitive Communications
\end{IEEEkeywords}

\section{Introduction}

A fundamental question for a communication system is when and how should two nodes communicate with each other? In traditional communication systems, this question is dealt with entirely by the human designers of the system, who pre-program the transmission and reception of messages, by explicitly accounting for different cases or scenarios. This works well for systems that will operate in relatively well-structured, deterministic and static environments. However, it is challenging for such systems to adapt their communication behavior when the  environment is more uncertain and dynamic. 

It is therefore of interest to develop cognitive communication systems that endow distributed agents with the autonomy and intelligence to learn when to communicate on their own, and the ability to re-learn the most efficient behavior given a change in environmental conditions.
Pursuing this goal, we define a game between two agents: a signaler and a responder. The signaler agent has a random, exogenous need and can choose from four different strategies: never signal, always signal, signal when need, and signal when no need. The responder agent can choose to either ignore or respond to the signal. We define a reward to both agents when they cooperate to satisfy the signaler's need, and costs associated with communication, response and unmet needs.

The optimal strategy for the signaler and responder will depend on these costs and rewards. Let's consider the strategy pair where the signaler signals whenever there is a need and the responder always responds. Intuitively, this could be an equilibrium for both players if the rewards are sufficiently higher than the cost for communication for the signaler and the cost of responding for the responder. However, if those conditions were to change and it becomes more expensive for the signaler to signal and for the responder to respond, there may be some other strategy that is preferable for both agents. We mathematically analyze and identify all the pure Nash equilibria for this game, and the conditions under which they occur. 

In practice, the agents need a distributed way to learn equilibrium strategies. For this we develop and present two Bayesian algorithms, one for the signaler and one for the responder, that are based on the well-known Thompson Sampling policy for multi-armed bandits \cite{ts1, ts2}. In these Bayesian algorithms, each agent updates a distribution for its belief about the random exogenous need, as well as the behavior of the other agent. Specifically, the signaler maintains and updates a distribution for its belief that a responder will respond to a signal. Likewise, the responder maintains and updates a distribution for its belief about whether the signaler has a need given that it signaled. Given a sample of their respective belief distributions, each agent selects a strategy which maximizes its expected reward. 

We designed a custom Python simulator to evaluate the performance of the distributed algorithms under different parameter settings pertaining to the reward and cost values that define the payoff matrix and initial belief distributions. We show through simulations that these distributed policies result in both agents converging to reward-maximizing equilibria of the signaler-responder game. Our simulation code is publicly available as an open-source at \url{https://github.com/anrgusc/l2com}

Further, we study the case where the rewards and costs for each agent are changing over time. In this case, we show that enhancing the policies to reset the belief distributions to a prior distribution when a change is detected, results in faster adaptation.

Our key contributions are the following:
\begin{itemize}
    \item We formulate the signaler-responder game which provides a simple framework to study the problem of two agents learning when to communicate and cooperate with each other.
    \item We analyze the pure Nash equilibria of this game.
    \item We present two new Bayesian distributed learning policies for both these agents using Thompson Sampling
    \item We design a custom simulation to evaluate these algorithms that has been made publicly available to ensure repeatability and to allow other researchers to build on our work. 
    \item We show through simulations that the proposed algorithms converge to efficient equilibria.
    \item For environments with changing rewards and costs, we show that it is important to detect the change and reset the beliefs to the initial distributions, to allow agents to adapt their strategies rapidly. 
\end{itemize}

The rest of the paper is organized as follows:
In section \ref{Sec:rw}, we discuss some relevant prior papers from the literature to place our work in context. In section \ref{Sec:game}, we define and present the strategies and payoffs for the signaler-responder game.  In section \ref{Sec:ne}, we analyze the Nash Equilibria of the proposed game. In section \ref{Sec:alg}, we present Bayesian distributed learning algorithms for each agent that are based on Thompson Sampling. In section \ref{Sec:sims}, we present simulations to evaluate the proposed algorithms and the equilibria they converge to. Finally, we present our concluding comments and thoughts regarding future work directions in section \ref{Sec:con}. 

\section{Related Work}
\label{Sec:rw}


The most closely related literature to our work is a 2012 paper titled ``How do communication systems emerge" by Scott-Phillips \emph{et al.}~\cite{sp}. This paper identifies two distinct pathways under which communication can emerge between two agents that they refer to as an actor and a reactor, respectively. The first pathway which they call ``ritualization" requires two conditions: (i) that there must be some action available to the actor that offers a positive benefit to the actor, \emph{independent} of any effect it may have on the reactor, and (ii) there is some reaction to it available to the reactor that offers them a positive benefit. In the signaler-responder game that we formulate, the first condition never holds true, because the only way for a signaler to earn a positive reward is if the responder responds to a signal when it has a need; thus our game does not satisfy this pathway. The second pathway, which they call ``Sensory manipulation" requires three conditions: (i) there must be some action available to the actor that (can) provoke a reaction from the reactor, (ii) the reaction should offer a positive reward to the reactor, and (iii) the reaction should offer a positive reward to the actor. The emergence of communication (when it occurs) in the signaler-responder game that we present in this paper is exactly consistent with this second pathway. Because we make "respond to signal" one of the strategies available to the responder, any strategies that result in the signaler sending a signal have the ability to provoke a response reaction. We observe through our analysis and simulations in this work that the agents converge to equilibria corresponding to the emergence of communication only if both the signaler and responder each get a net positive reward from signaling and responding to the signal, respectively. 


A significant domain of research that is quite close to our work and that has a large and growing literature is that of multi-agent reinforcement learning (MARL)~\cite{marl1, marl2}. In MARL, typically, multiple agents are placed in an environment where they must learn to communicate and cooperate to complete some task efficiently~\cite{marl2}. Examples of applications of MARL have been to the games of Starcraft~\cite{marlstarcraft} predator and prey~\cite{marlpp1, marlpp2}. These works consider highly complex environments and use large data driven algorithms such as deep reinforcement learning. While in these works the MARL algorithms are empirically shown to do well both in terms of cooperative task performance and in terms of learning efficient communications, because of the complexity of the environment and the algorithms themselves, it can be hard to derive clean and generalizable mathematical insights about when communication emerges. In contrast, the game we propose and analyze is intentionally simple, with only two agents, a total of 8 strategy pairs, and closed-form expressions for the payoff to agents under different strategy pairs and environmental conditions. This allows us to mathematically analyze the equilibria and identify the conditions under which communication emerges or not. We believe that focusing on a more simplified setting yields general insights and argue that this methodology nicely complements the more complex studies involving MARL when it comes to understanding how agents can learn to communicate efficiently without being pre-programmed to do so.


In order to come up with a distributed learning algorithm for each of the players we utilize a Bayesian method known as Thompson Sampling. Thompson Sampling is a very well known algorithm for finding reward maximizing strategies in stochastic multi-armed bandit problems~\cite{ts1, ts2}. It has been applied previously to the problem of learning in stochastic games\cite{tsg1, tsg2}. DiGiovanni and Tiwari~\cite{tsg2} have shown that incorporating change detection and resetting the beliefs to a prior distribution is helpful to improve the adaptation of Thompson Sampling in Markov Games to dynamic environments, which is consistent with the observation we also make in this work. 

\section{The Signaler-Responder Game}
\label{Sec:game}

We formulate the problem as a game between two agents, a signaler and a responder. There is an exogenous ``need" that is generated at the signaler end, drawn from a Bernoulli distribution with probability $p_{n}$. The signaler can choose between four strategies: s0 -- never signal; s1 -- always signal; s2 -- signal only when there is a need; and s3 -- signal when there is no need, don't signal when there is a need. The responder then has a choice to respond or not to the signal, resulting in the following two strategies: r0 -- ignore signals; r1 -- respond to signals.

We model the following costs and rewards:  let $R$ be a reward that is given to both the signaler and responder when three conditions are met: (a) there is a need, (b) the signaler signals, and (c) the responder responds to address the need; let $\rho_{um}$ be the cost associated with an unmet need, i.e. when the signaler has a need but either the signaler does not signal or it signals but the responder does not respond; let $\rho_{t}$ be the trip penalty, a cost for the responder to respond to a signal; and let $\rho_{com}$ be the communication cost for the signaler to signal. Given these rewards and costs, we can write down the payoff matrix for the game - see table~\ref{tab:payoff}.

\begin{table*}[hbt]
\vspace{.1in}
\centering
\begin{tabular}{|c|c|c|}
\hline
 & \textbf{r0} & \textbf{r1} \\ 
\hline
\textbf{s0} & $(-p_n\cdot \rho_{um},0)$ & $(-p_n\cdot\rho_{um},0)$ \\\hline
\textbf{s1} & $(-\rho_{com}-p_n\cdot \rho_{um},0)$ & $-\rho_{com}+p_n\cdot \rho_{um}p_n\cdot R -\rho_{t})$ \\ \hline
\textbf{s2} & $(-\rho_{com}\cdot p_n -p_n\cdot \rho_{um},0)$ & $(-\rho_{com}\cdot p_n +p_n\cdot\ R,p_n\cdot R -p_n\cdot \rho_{t})$ \\ \hline
\textbf{s3} & $(-\rho_{com}(1-p_n)-p_n\cdot \rho_{um},0)$ & $(-\rho_{com}(1-p_n)-p_n\cdot\rho_{um},-(1-p_n)\cdot \rho_{t})$ \\ 
\hline
\end{tabular}
\caption{\label{tab:payoff}The Signaler Responder Game}
\end{table*}

\section{Nash Equilibrium Analysis}
\label{Sec:ne}

Given the payoff matrix for the game defined in table~\ref{tab:payoff}, we can derive conditions under which various strategy pairs form pure Nash equilibria. For brevity, we omit the details of the derivation of these conditions in this paper, but they are straightforward to derive from the definition of a pure Nash equilibrium, that neither agent has an incentive to deviate from the given strategy. These are as given in table~\ref{tab:nash}. 

It is important to note that these are not the only possible Nash equilibria for the given game. There can also be mixed Nash equilibria that are obtained by the agents randomizing between pure strategies. In a subsequent section we will see that the distributed learning algorithms we propose sometimes result in a mixed Nash equilibrium. 

One notable observation is that the trivial strategy pair of $(s0, r0)$, where the signaler never signals and the responder never responds, is always a Nash equilibrium of the game. Intuitively, it would be desirable for the agents to adopt a learning strategy for this game that tries to avoid this trivial outcome when there are other equilibria that offer better rewards to the agents.  

\begin{table*}[hbt]
\vspace{.1in}
\centering
\begin{tabular}{|c|c|}
\hline
 \textbf{Pure Strategy Pair} & \textbf{Conditions under which it is a Nash equilibrium} \\ 
\hline
(s0,r0) & always a \emph{NE} \\
\hline
(s0,r1) & $\rho_{com}\geq R+\rho_{um}$ \\ 
\hline
(s1,r0) & $\rho_{com} = 0\: and\: p_n\cdot R - \rho_{t} \leq 0$ \\ 
\hline
(s1,r1) & $\rho_{com} = 0\: and\: p_n\cdot R - \rho_{t} \geq 0$ \\ 
\hline
(s2,r0) & $\rho_{com} = 0\: and\: p_n\cdot R - p_n\cdot \rho_{t} \leq 0$ \\
\hline
(s2,r1) & $R \geq \rho_{com} - \rho_{um}\: and\: R \geq \rho_{t}$ \\
\hline
(s3,r0) & $\rho_{com}=0$ \\ 
\hline
(s3,r1) & Never a \emph{NE} \\ 
\hline
\end{tabular}
\caption{\label{tab:nash} Conditions for pure Nash equilibria}
\end{table*}

\section{Thompson Sampling-based Learning Algorithm}
\label{Sec:alg}



\subsection{Learning from the Signaler's Perspective}
Let us define $\theta_A$ as the probability that the signaler has a need. This is a Bernoulli event, and to simplify the Bayesian update, we will model the prior distribution of $\theta_A$ as $Beta(\alpha_A, \beta_A)$. We assume that the event that the responder responds when a signaler has sent out a signal can also be modeled as a Bernoulli event, with probability $\theta_B$. The prior distribution for $\theta_B$ is likewise modeled as $Beta(\alpha_B, \beta_B)$ for convenience. We assume that these parameters $\alpha_A, \beta_A, \alpha_B, \beta_B$ are initially set to some default values, e.g., we can set them all to 1 initially. For each strategy, we can write an expression for the expected reward given the value of the probabilities $\theta_A$ and $\theta_B$, as follows:

\begin{itemize}
    \item $E[R_{s,0}(\theta_A, \theta_B)] = -\rho_{UM} \theta_A $
    \item $E[R_{s,1}(\theta_A, \theta_B)] = R\theta_A\theta_B - \rho_{UM}\theta_A(1-\theta_B) - \rho_{COM}$
    \item $E[R_{s,2}(\theta_A, \theta_B)] = R\theta_A\theta_B - \rho_{UM}\theta_A(1-\theta_B) - \rho_{COM}\theta_A$
    \item $E[R_{s,3}(\theta_A, \theta_B)] = -\rho_{UM} \theta_A  - \rho_{COM}(1-\theta_A)$
\end{itemize}

\begin{algorithm}
\caption{Thompson Sampling-based Strategy Selection for Signaler}\label{alg:signaler}
\begin{algorithmic}
\For{\texttt{$t = 1, 2, \ldots T$}}
    \State \textbf{Strategy Selection Phase}
    \State {Sample $\theta_A$ from $Beta(\alpha_A, \beta_A)$ }
    \State {Sample $\theta_B$ from  $Beta(\alpha_B, \beta_B)$ }
    \State {Compute $E[R_{s,i}(\theta_A, \theta_B)]~ \forall i \in \{0, 1, 2, 3\}$ } 
    \State {Play strategy $i(t) := arg max_i E[R_{s,i} (\theta_A, \theta_B)]$  and observe the outcomes}
    \State \textbf{Belief Update Phase}
    \State if need observed  $\alpha_A \pluseq 1$, else $\beta_A += 1$
    \State if signal sent and responder responded $\alpha_B += 1$ ; if signal sent and responder did not respond $\beta_B += 1$
\EndFor
\end{algorithmic}
\end{algorithm}

\subsection{Learning from the Responder's Perspective}
We model the event that the signaler has a need given that it has signaled as a Bernoulli event with probability $\theta_C$. To simplify the Bayesian update, we will model the prior distribution of $\theta_C$ as $Beta(\alpha_C, \beta_C)$. For each strategy, we can then write an expression for the expected reward given the value of $\theta_C$ as follows:

\begin{itemize}
    \item $E[R_{r,0}(\theta_C)] = 0 $
    \item $E[R_{r,1}(\theta_C)] = R\theta_C - \rho_{T}$
\end{itemize}

 \begin{algorithm}
 \caption{Thompson Sampling-based Strategy Selection for Responder}\label{alg:responder}
\begin{algorithmic}
\For{\texttt{$t = 1, 2, \ldots T$}}
    \State \textbf{Strategy Selection Phase}
    \State {Sample $\theta_C$ from $Beta(\alpha_C, \beta_C)$ }
    \State {Compute $E[R_{r,i}(\theta_C)]~ \forall i \in \{0, 1\}$ } 
    \State {Play strategy $i(t) := arg max_i E[R_{r,i} (\theta_C]$  and observe the outcomes}
    \State \textbf{Belief Update Phase}
    \State if the signaler signaled and the responder responded $\alpha_C += 1$, else $\beta_C += 1$
\EndFor
\end{algorithmic}
\end{algorithm}

\subsection{Distributed Iteration}
Given the algorithms above, the two agents play the game repeatedly as follows. At each iteration, they each first select a strategy based on their beliefs, play that strategy and collect observations in  a distributed manner. Now the signaler observes if it has a need, and if it signals it can observe if the responder responded. This suffices for the signaler to update its beliefs about $\theta_A$ and $\theta_B$. Likewise, the responder observes if the signaler has a need or not given that the signaler signaled and it responded. This enables it to update its belief about $\theta_C$ accordingly. 


\section{Simulation Results}
\label{Sec:sims}

In this section we present the results of our simulations, that identify the conditions under which equilibria are reached by the learning algorithms.

By default, we assume the following parameter values in our simulations unless otherwise specified. $R = 1, \rho_t = 0.8, \rho_{com} = 0.5, \rho_{um} = 0.5, p_n = 0.8$, and initially, we set $\alpha_A = \alpha_B = \alpha_C = \beta_A = \beta_B = \beta_C = 2$. We run the learning for each case for a maximum of 20000 iterations as a default.

\subsection{Signaling when need and always responding}
\begin{figure}[htbp]
\centerline{\includegraphics[scale=0.4]{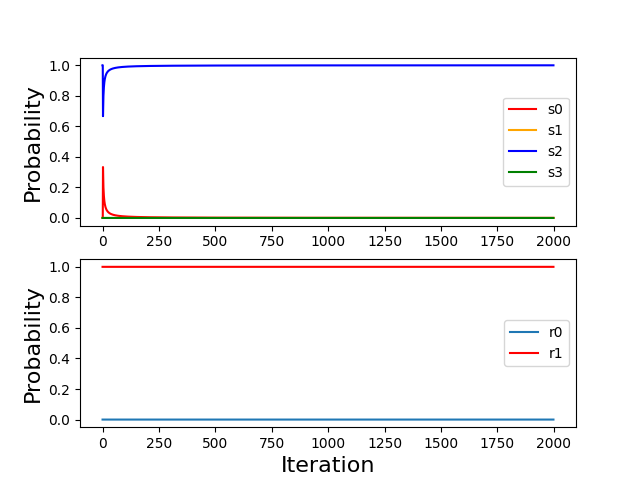}}
\caption{Strategies learned by signaler and responder when  $R = 1, \rho_t = 0.8, \rho_{com} = 0.5, \rho_{um} = 0.5, p_n = 0.8$ }
\label{fig:s2r1}
\end{figure}

First, we see the result under the default parameters indicated above; in this case, as we can see in figure~\ref{fig:s2r1}, we can see the reward is sufficiently high both for the signaler to signal and for the responder to respond and the learning converges to this desirable outcome. 

\subsection{Failing to Communicate}
\begin{figure}[htbp]
\centerline{\includegraphics[scale=0.4]{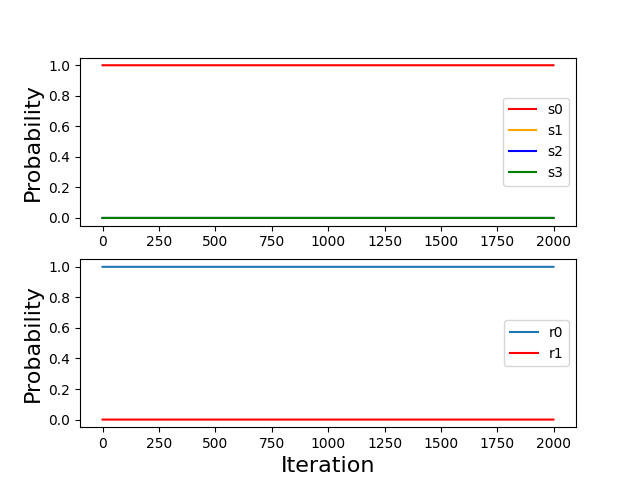}}
\caption{Strategies learned by signaler and responder when  $R = 1, \rho_t = 2, \rho_{com} = 2, \rho_{um} = 0.5, p_n = 0.8$}
\label{fig:s0r0}
\end{figure}

Next, we consider a setting where the signaler should learn to not communicate at all (i.e., adopt strategy $s0$) and the responder in turn to never respond (i.e., adopts strategy $r0$). While our analysis shows that this is always an equilibrium, we expect it particularly to be the only equilibrium when the communication cost is high for the signaler $\rho_{com} > R + \rho_{um}$ and when the trip penalty is high for the receiver $\rho_{t} > R$. Such a scenario could be observed in figure~\ref{fig:s0r0} where we set $\rho_{com} = \rho_t = 2$, while keeping all other values at the default setting. 

\subsection{Zero comm-cost}

Next, we consider some settings where the signaler incurs no cost for communication. Intuitively, we may expect to see the signaler choose to always signal at least some of the time, since there is no cost associated with this strategy. However the response of the responder would depend on its trip cost and whether it gets sufficiently rewarded due to need of the signaler. 

\begin{figure}[htbp]
\centerline{\includegraphics[scale=0.4]{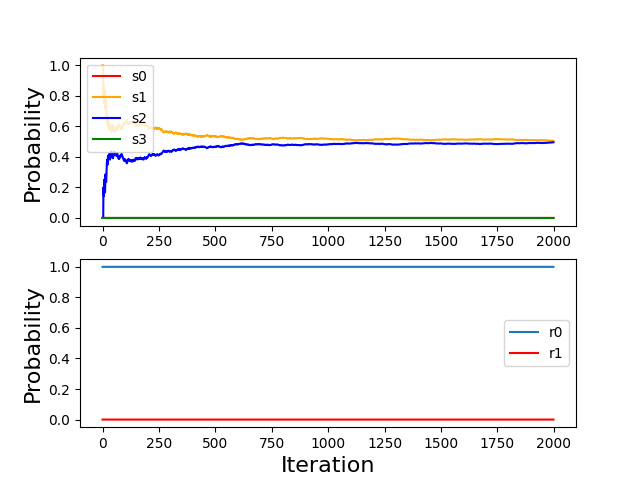}}
\caption{Strategies learned by signaler and responder when  $R = 1, \rho_t = 2, \rho_{com} = 0, \rho_{um} = 0.5, p_n = 0.8$}
\label{fig:s12r0}
\end{figure}

\textbf{high trip penalty case:} As seen in figure~\ref{fig:s12r0}, when the responder has a high trip penalty compared to reward, while communication is free for the signaler, we observe a mixed equilibrium where the signaler mixes evenly between always signaling and signaling only when it has a need, but the responder does not respond. 

\begin{figure}[htbp]
\centerline{\includegraphics[scale=0.4]{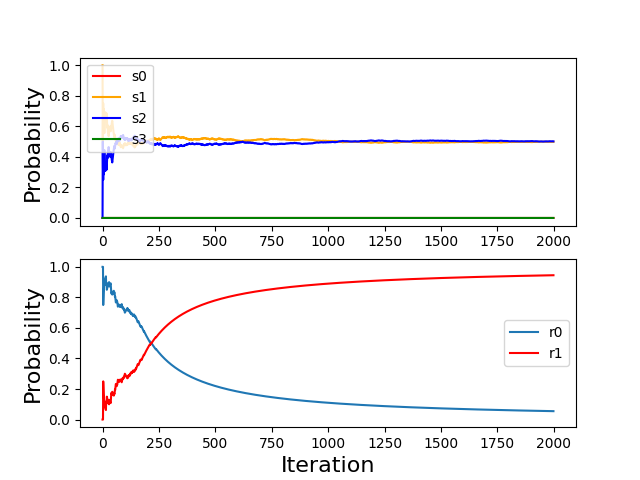}}
\caption{Strategies learned by signaler and responder when  $R = 1, \rho_t = 0.8, \rho_{com} = 0, \rho_{um} = 0.5, p_n = 0.8$}
\label{fig:s12r1}
\end{figure}

\textbf{low trip penalty but high need:}  We consider the case when the cost for signaling is 0 and the responder incurs a low  trip penalty compared to reward and the probability of need is relatively high. In this case the responder finds its response pays off most of the time. As seen in figure~\ref{fig:s12r1}, we observe that this results in a mixed equilibrium where the signaler mixes evenly between always signaling and signaling only when it has a need, and the responder always responds. 

\textbf{low trip penalty but low-need:} A different scenario that merits investigation is when there is zero communication cost, and the trip penalty of the responder is less than the reward, however, the need of the signaler is low. In this case, the responder may not be rewarded most of the time when it makes the trip, resulting in a negative expected reward from responding. For such a setting (specifically,  $R = 1, \rho_t = 0.8, \rho_{com} = 0, \rho_{um} = 0.5, p_n = 0.1$) we observe that the algorithms converge to an equilibrium where the responder does not respond, though the signaler mixes equally between the strategies of always signaling and signaling when it has a need (we omit the figure for this case due to space constraints). 




\subsection{Summary of findings with static rewards}
Based on the experiments described above, all involving rewards and penalties that do not change over the course of the simulation, we observe the following: 
\begin{itemize}
\item The proposed Thompson sampling based distributed learning policy always converges to a single pure or mixed Nash equilibrium. 
\item When there are multiple possible equilibria, the strategies converge to efficient equilibria; this can be attributed to the utility-maximizing nature of Thompson Sampling. Because we split ties randomly in both algorithms, the algorithms converge to mixed strategy equilibria with equal probabilities whenever there are multiple pure strategies offering the same net reward.
\item In our simulations, we never observe convergence to $s0, r1$ even in conditions where this is a (trivial) equilibrium (such as $\rho_{t} = 0$); this is because the responder doesn't get an opportunity to learn how this strategy would perform in the absence of any signals. 
\end{itemize}

\subsection{Time-varying rewards and costs}
\begin{figure}[htbp]
\centerline{\includegraphics[scale=0.4]{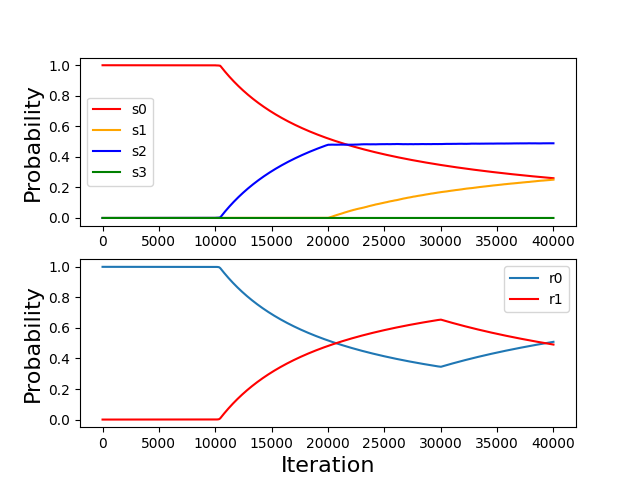}}
\caption{Strategies learned by signaler and responder in the presence of time-varying rewards and costs without belief reset}
\label{fig:nochange}
\end{figure}

\begin{figure}[htbp]
\centerline{\includegraphics[scale=0.4]{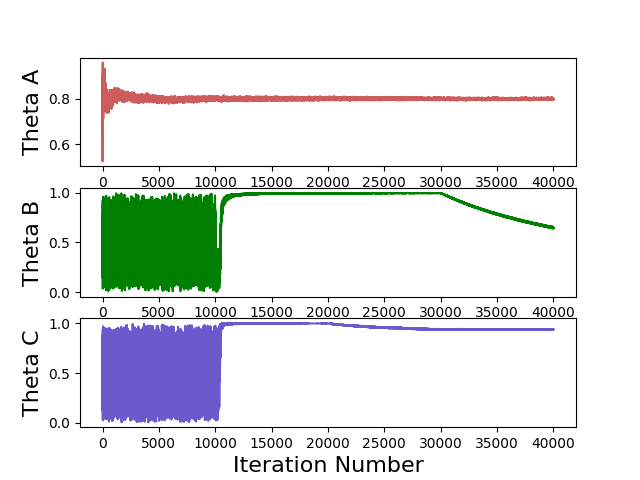}}
\caption{Belief distributions for $\theta_A, \theta_B, \theta_C$ in the presence of time-varying rewards and costs without belief reset}
\label{fig:thetanochange}
\end{figure}

\begin{figure}[htbp]
\centerline{\includegraphics[scale=0.4]{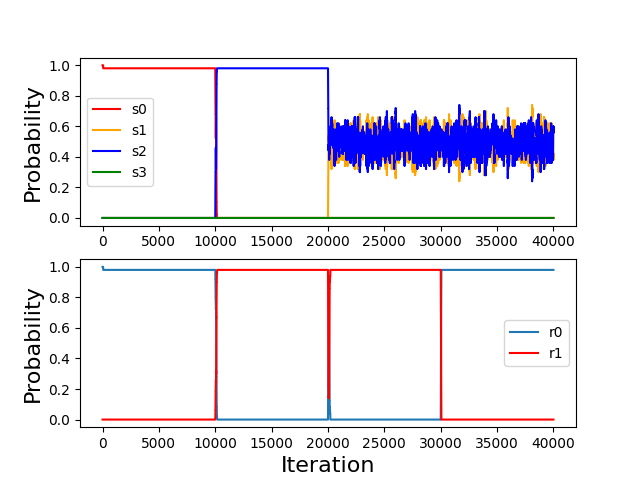}}
\caption{Strategies learned by signaler and responder in the presence of time-varying rewards and costs using belief reset}
\label{fig:change}
\end{figure}

\begin{figure}[htbp]
\centerline{\includegraphics[scale=0.4]{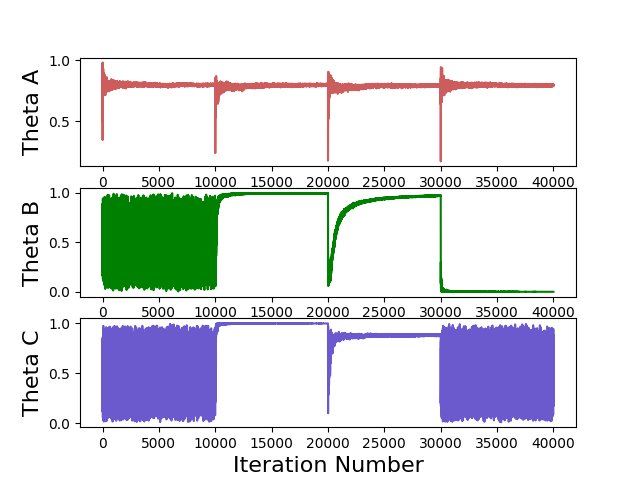}}
\caption{Belief distributions for $\theta_A, \theta_B, \theta_C$ in the presence of time-varying rewards and costs using belief reset}
\label{fig:thetachange}
\end{figure}

We do an additional set of simulations where we change the rewards and costs at different times. Specifically, we choose the following reward and cost values in these simulations: till iteration 10000, we use the default parameters; from iterations 10001-20000, ($R = 1, \rho_t = 0.8,\rho_{com} = 0.5, \rho_{um} = 0.5, p_n = 0.8)$; from iterations 20001-30000, ($R = 1, \rho_t = 0.8 ,\rho_{com} = 0.5 , \rho_{um} = 0.8, p_n = 0.8)$; and from iterations 30001-40000, ($R = 1, \rho_t = 2,\rho_{com} = 0, \rho_{um} = 0.5 p_n = 0.8)$. We observe in figure~\ref{fig:nochange} that the agents start to change their strategies whenever the conditions change, moving towards new efficient equilibria. However, the adaptation is quite slow. We observe that this is because by default, after each change the agents both have to first ``unlearn" their strong beliefs learned during the previous environment, resulting in slow changes to the belief distribution, as seen in figure~\ref{fig:thetanochange}. To speed up the process of adaptation therefore, we recommend resetting the beliefs to the initial prior distribution after each change. We see the significantly improved performance of this approach in terms of rapid convergence to new strategies after each change in figure~\ref{fig:change} and rapid convergence of beliefs to new values after each change in figure~\ref{fig:thetachange}. 
We should note that while we have assumed in our work, for simplicity, an oracle that provides knowledge of the changed environmental condition to both agents to implement belief reset instantaneously, it is possible to explicitly incorporate a change detection mechanism for each agent, as proposed in~\cite{tsg2}.

\section{Conclusions}
\label{Sec:con}

We have presented the signaler-responder game, a simple but rich model which has allowed us to mathematically explore conditions under which heterogeneous agents can learn to communicate and cooperate with each other. We have proposed distributed Bayesian learning algorithms based on Thompson Sampling that the agents can adopt. We have shown through simulations that these algorithms converge to reward-maximizing equilibria for the agents, and with some additional modification to reset initial beliefs, could be made adaptive to changing conditions. 

We have assumed in this work that the various costs and rewards in the game are deterministic and known to each agent who use them when calculating the reward-maximizing strategy at each iteration. In future work, these rewards and costs could be assumed to be stochastic and observed and estimated in an online fashion 
We believe that this will not fundamentally change the conclusions, although the learning algorithms may take longer to converge in the presence of uncertainty in the rewards and penalties. In ongoing work, we are also exploring extending the simple two-player formulation in this work to consider many sets of signaler and responder agents embodied in the form of distributed robots moving in a given environment.

\end{document}